\documentclass[conference]{IEEEtran}
\IEEEoverridecommandlockouts
\usepackage{graphicx}
\usepackage{float}
\usepackage{stackengine}
\usepackage{tabularx}
\usepackage{array}
\usepackage{dirtytalk}
\usepackage{caption}
\usepackage{cite}
\usepackage{amsmath,amssymb,amsfonts}
\usepackage{algorithmic}
\usepackage[ruled,lined]{algorithm2e}
\usepackage{textcomp}
\usepackage{xcolor}
\usepackage{dblfloatfix}
\usepackage{balance}
\usepackage{breqn}
\newcolumntype{Y}{>{\centering\arraybackslash}X}

\def\BibTeX{{\rm B\kern-.05em{\sc i\kern-.025em b}\kern-.08em
    T\kern-.1667em\lower.7ex\hbox{E}\kern-.125emX}}
\begin{document}

\title{ADORN: Adaptive Drift handling for Open RAN using Reinforcement Learning}

\author{\IEEEauthorblockN{Ashit Kumar Subudhi\IEEEauthorrefmark{1},
Bhargav Chirumamilla\IEEEauthorrefmark{3}, Shubham Vaishnav\IEEEauthorrefmark{5}, Mduduzi C. Hlophe\IEEEauthorrefmark{2}, \\ Praveen Kumar Donta\IEEEauthorrefmark{5},  Andrea Fumagalli\IEEEauthorrefmark{4}, Venkateswarlu Gudepu\IEEEauthorrefmark{4}, Koteswararao Kondepu\IEEEauthorrefmark{1}
}
\IEEEauthorblockA{\IEEEauthorrefmark{1} Indian Institute of Technology Dharwad, Karnataka, India\\ 
\IEEEauthorrefmark{2} University of Pretoria, Pretoria, South Africa
\IEEEauthorrefmark{3} Johns Hopkins University, USA \IEEEauthorrefmark{5} Stockholm University, Sweden \\
\IEEEauthorrefmark{4}Open Networking Advanced Research (OpNeAR) Lab, The University of Texas at Dallas, TX, USA\\
Email: ashit.subudhi.21@iitdh.ac.in}
}

\maketitle

\begin{abstract}
Dynamic traffic variations in Open Radio Access Networks (O-RAN) lead to drift, which degrades the performance of Artificial Intelligence/Machine Learning (AI/ML) models.
Traditional retraining approaches maintain forecasting accuracy but incur high computational cost and may lead to violations of Service Level Agreements (SLAs).
This work proposes a Q-learning-based adaptive retraining approach that formulates the retraining decision as a Markov Decision Process (MDP), where a Reinforcement Learning (RL) agent learns a policy that balances forecasting accuracy and retraining cost.
The proposed approach incorporates a multi-expert Long Short-Term Memory (LSTM) ensemble to mitigate catastrophic forgetting and improve robustness across diverse traffic conditions.
Experimental results show that the proposed approach effectively reduces retraining overhead compared to greedy and random baselines, while maintaining system performance within predefined limits.

\end{abstract}


\section{Introduction}

The advent of fifth-generation and beyond (B5G) networks marks a significant leap forward in telecommunications and enables a wide range of new services --- network slicing, autonomous vehicles, Augmented and Virtual Reality (AR/VR), and e-Health. 
The B5G networks support high data rates, ultra-low latency, and connectivity for a massive number of devices, which are critical requirements for addressing diverse and dynamic user service demands.
However, traditional Radio Access Network (RAN) architectures rely on proprietary hardware with closed and embedded interfaces, which limits flexibility and restricts configurability to meet B5G network requirements~\cite{polese2023understanding}.

The O-RAN Alliance introduces RAN Intelligent Controllers (RICs) that enable data collection across the RAN through standardized open interfaces and incorporate intelligence using Artificial Intelligence and Machine Learning (AI/ML) algorithms.
However, AI/ML model performance relies on the characteristics of the training data. 
The arrival of new and previously unseen user traffic patterns introduces user data traffic changes, which degrade AI/ML model performance, referred to as drift. 
Drift caused by dynamic fluctuations in user traffic produces false insights and inaccurate decisions, leading to poor network management, severe service interruptions, inefficient resource allocation, and overall network performance degradation. 
Such effects result in violations of Service-Level Agreements (SLAs)~\cite{gudepu2024drift}, along with reduced network reliability, increased operational costs, and degraded user experience.
Such consequences can be catastrophic for time-critical applications --- smart ambulances --- where drift-induced performance degradation leads to failures in meeting stringent throughput and response-time requirements.

Drift handling for B5G networks primarily relies on performance monitoring, statistical analysis, and data-driven approaches.
Threshold-based approaches detect drift when AI/ML model performance metrics --- accuracy, precision, recall, or error rate --- exceed predefined thresholds, with traditional approaches including Drift Detection Method (DDM) and Early Drift Detection Method (EDDM)~\cite{lu2018learning}. 
Statistical approaches such as the Fisher score evaluate changes in feature user traffic to identify drift without relying on model error rates~\cite{zhang2020concept}. 
Existing approaches rely on predefined thresholds, require large data samples, and struggle to adapt to highly dynamic, complex network environments. 
Selecting appropriate threshold values remains challenging: low thresholds increase computational overhead due to frequent retraining, while high thresholds degrade model performance and lead to SLA violations. 
Data stream-based approaches also suffer from limited adaptability, as static or periodic evaluations fail to capture the continuously evolving nature of real-world data.

In~\cite{3152494}, the authors introduced a drift adaptation and anomaly detection algorithm for Recurrent Neural Networks (RNNs), where the prediction model is updated incrementally without directly detecting drift. 
However, the adoption of Reinforcement learning (RL)-based approaches enables adaptive decision-making by learning when to trigger model retraining based on observed system states and has shown potential to improve the trade-off between prediction performance and retraining frequency in dynamic environments.

The RL improves threshold-based retraining approaches by enabling adaptive decision-making, rather than static retraining rules, when drift is detected. 
The RL aims to maximize the cumulative reward, while making proactive and adaptive retraining decisions, unlike conventional static decision-making. 
Whereas the proposed RL-based approach learns an adaptive retraining policy that dynamically decides when to update the model based on observed system states.
Q-learning, a model-free RL approach, is well-suited to the uncertain and dynamic nature of B5G systems, as it does not rely on a system model. In addition, it enables effective exploration-exploitation balances, enabling the proposed approach to dynamically adapt to different traffic patterns while ensuring reliable performance~\cite{liu2023leaf}.

To overcome the shortcomings of threshold-based decision-making schemes combined with traditional predictive or unsupervised methods for model retraining, this paper proposes an RL-based approach for drift handling in B5G networks. 
The main contributions of the proposed ADORN approach are summarized as follows: 
\begin{itemize}
    \item An RL-based Q-learning approach for drift handling.
    \item A state representation using statistical features of user traffic.
    \item An adaptive Long Short-Term Memory (LSTM) ensemble for dynamic model selection.
    \item An evaluation using Quality of Service (QoS) prediction use case with real-time and synthetic datasets.
    \item A performance comparison of the proposed work with Greedy and Random baseline approaches.
\end{itemize}

\begin{figure*}[!htb]
\centering
\vspace{-0.4cm}
\includegraphics[width=\linewidth]{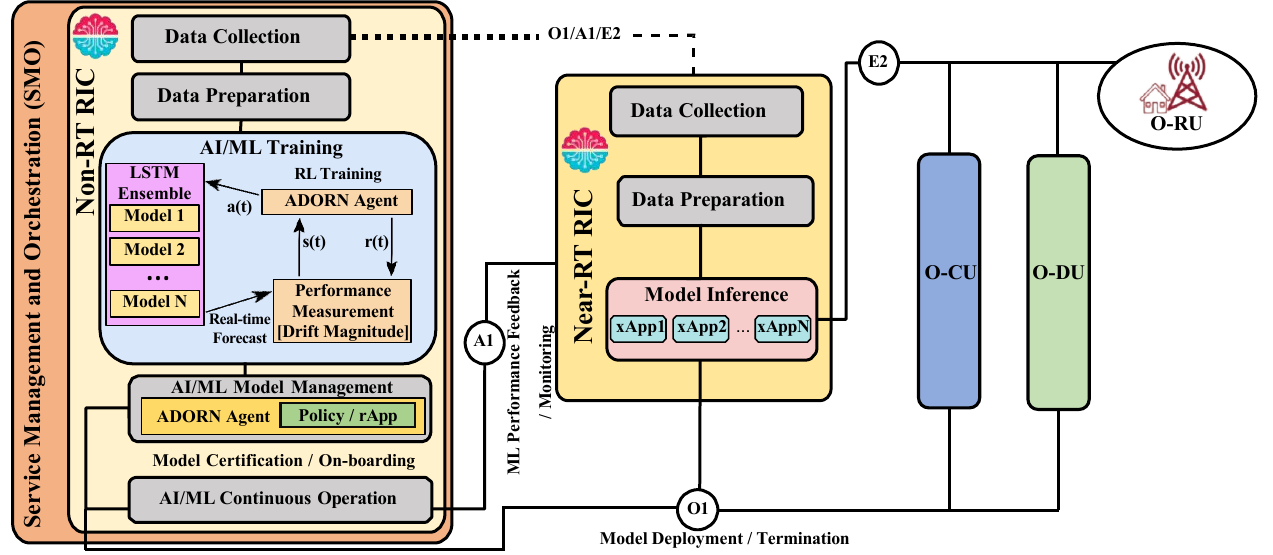}
\caption{System Model}
\label{fig:arch_ants24}
\vspace{-0.7cm}
\end{figure*}

\section{System Model and RL Problem Formulation}
\vspace{0.3cm}
\label{system}
The proposed system model is based on the O-RAN architecture, defining RICs, i.e., the Near-Real Time (Near-RT RIC) and the Non-Real Time (Non-RT RIC), which enables the \emph{intelligence} as illustrated in Fig.~\ref{fig:arch_ants24}.
The RICs enable autonomous optimization of O-RAN by functioning at different timescales, depending on the AI/ML model inference position.
For instance, the Non-RT RIC corresponds to the operations that has granularity of $>1\ sec$ --- Offline AI/ML or RL agent training, policy creation and enforcement, and many other; whereas the Near-RT RIC handles functionalities --- load balancing, handover, scheduling policy, RAN slicing and many other --- that has time scale between $10\ ms$ and $<1\ sec$.
Inside the Non-RT RIC, the AI/ML models or RL agents are trained and deployed as rApps. 

The O-RAN architecture provides multiple interfaces --- O1, A1, and E2 ---, to facilitate data collection and communication among the RAN components (i.e., central unit (O-CU), distributed unit (O-DU), and radio unit (O-RU)). 
The AI/ML model management block within the Non-RT RIC plays a key role in drift handling utilizing the proposed RL-based approach, which leverages an RL agent.
The trained RL agent generates optimal policies to determine the drift occurrence and communicates with the AI/ML model management block whenever drift is detected.
The RL agent can also be deployed as an rApp to further enhance its integration within the RIC.

To address drift in B5G networks due to dynamic changes in user traffic, the proposed work employs an ensemble of LSTM models that continuously evaluate performance under evolving network conditions.
The proposed approach formulates the retraining decision as a sequential decision-making problem using a Markov Decision Process (MDP) and trains an RL agent to select retraining actions based on observed system states, with the objective of maximizing predictive accuracy while minimizing computational cost. 

The proposed approach feeds normalized Mean Absolute Error (nMAE) along with statistical features into the RL agent to distinguish normal variations from drift and to balance performance degradation with the number of retraining actions, and the proposed approach defines the RL problem using state, action, and reward components.

\subsection{Preliminaries}

The proposed approach defines the problem as a MDP represented by the tuple $(\mathcal{S}, \mathcal{A}, \mathcal{P}, R, \gamma)$, where $\mathcal{S}$ denotes the state space, $\mathcal{A}$ denotes the action space, $\mathcal{P}(s' \mid s, a)$ defines the state transition dynamics, $R(s, a)$ represents the reward function, and $\gamma$ is the discount factor.
The elements of the MDP are defined as follows:

\subsubsection{State Space}
The state space is based on statistical characteristics of incoming traffic data, and the framework represents each state as a discretized 2-tuple:
\begin{equation}
{\bf s_t} \triangleq \left(\textnormal{mean}, \textnormal{variance}\right) \in \mathcal{S} \label{eq:state}
\end{equation}
where $\mathcal{S}$ denotes the set of all possible states, and $s_t$ denotes the system state at time $t$.

\subsubsection{Action Space}
The action space is defined as a binary decision to optimize the retraining policy of the specialized LSTM ensemble. At each discrete decision point, the reinforcement learning agent selects an action $a_t$ as follows:
\begin{equation}
{\bf a_t} \triangleq \begin{cases}
1,\quad &\mbox{$\textnormal{Retrain the model}$} \\
0, \quad &\mbox{$\textnormal{Remain the same (Idle)}$}
\end{cases}\label{eq:action}
\end{equation}
where $a_t=1$ triggers immediate model retraining to restore predictive accuracy upon drift detection, and $a_t=0$ avoids unnecessary retraining and preserves computational resources.

\subsubsection{Reward Function}
The reward function is $R(s, a): \mathcal{S} \times \mathcal{A} \rightarrow \mathbb{R}$ to balance predictive accuracy and computational cost.
The reward function incorporates sensitivity to both the presence and magnitude of drift through a magnitude-aware piecewise formulation.
Let $e_t$ denote the nMAE of a particular traffic window, and let $\theta$ denote the predefined drift threshold.
The drift magnitude as $\Delta_t = e_t - \theta$. The reward $R_t$ for an action $a_t \in {0, 1}$ is given by:
\begin{equation}
R_t(s_t, a_t) =
\begin{cases}
w_{1}|\Delta_t| + C_{1}, & \text{if } a_t=1, \ \Delta_t > 0 \\
w_{2}|\Delta_t| + C_{2}, & \text{if } a_t=0, \ \Delta_t \le 0 \\
-w_{3}|\Delta_t| - C_{3}, & \text{if } a_t=1, \ \Delta_t \le 0 \\
-w_{4}|\Delta_t| - C_{4}, & \text{if } a_t=0, \ \Delta_t > 0
\end{cases}
\end{equation}
where $w_{1}, w_{2}, w_{3}, w_{4}$ denote weighting coefficients that control sensitivity to drift magnitude, and $C_{1}, C_{2}, C_{3}, C_{4}$ denote operational constants that represent fixed rewards or penalties associated with each decision.

\subsection{Objective Function and Learning Strategy}

The proposed approach aims to achieve a balance between high predictive accuracy and low cumulative retraining actions by maximizing the expected cumulative discounted reward:
\begin{equation}
J(\pi) = \mathbb{E}{\tau \sim \pi{\theta}} \left[ \sum_{t=0}^{\infty} \gamma^t r(s_t, a_t) \right]
\end{equation}
where $\tau = (s_0, a_0, r_0, s_1, a_1, \dots)$ denotes the trajectory of states, actions, and rewards generated under policy $\pi_{\theta}$, $\mathbb{E}$ denotes the expectation over traffic scenarios, and $\gamma \in [0,1]$ denotes the discount factor.
The proposed approach employs a Q-learning algorithm~\cite{sutton2018reinforcement} to learn an optimal policy through interaction with the environment. 
The Q-learning algorithm updates the action-value function $Q(s_t,a_t)$ using the Bellman update rule:
\begin{equation}
Q(s_t,a_t) \leftarrow Q(s_t,a_t) + \alpha \left[ r_t + \gamma \max_{a'} Q(s',a') - Q(s_t,a_t) \right]
\end{equation}
where, $\alpha$ denotes the learning rate, $s'$ denotes the next state after taking action $a_t$, $a'$ denotes the set of possible actions in state $s'$, and $\max_{a'} Q(s', a')$ denotes the maximum expected future reward.
This learning process enables the agent to learn a policy that triggers retraining only when the long-term gain in prediction accuracy outweighs the associated computational cost, thereby ensuring efficient and reliable operation in dynamic B5G networks.

\begin{table}[!htb]
\centering
\caption{Summary of Key Notations and Symbols}
\label{tab:nomenclature}
\begin{tabular}{ll}
\hline
\textbf{Symbol} & \textbf{Description} \\ \hline
$W$ & Incoming 5G Traffic Stream (Temporal Window) \\
$\{\mathcal{E}_i\}$ & Ensemble of Specialized LSTM Models \\
$\mathcal{S}$ & State Space (discretized traffic statistics) \\
$a_t$ & Action ($a_t=0$: Idle, $a_t=1$: Retrain) \\
$e_t$ & Real-time Forecast Error (nMAE) \\
$\theta$ & Drift Allowance Threshold (SLA Boundary) \\
$\Delta_t$ & Drift Magnitude ($e_t - \theta$) \\
$R_t(s,a)$ & Instantaneous Reward Function \\
$Q(s,a)$ & State-action value function (Q-value) \\
$\alpha$ & Learning rate \\
$\gamma$ & Discount factor \\
$\epsilon$ & Exploration probability for the $\epsilon$-greedy policy \\
$w_i, C_i$ & Scaling weights and constant offsets for the reward function \\
$\pi^*$ & Optimized Drift Orchestration Policy \\ \hline
\end{tabular}
\end{table}

\begin{algorithm}
\caption{RL-Based Adaptive Drift Handling}
\label{alg:rl_drift}
\begin{algorithmic}[1]
\renewcommand{\algorithmicrequire}{\textbf{Input:}}
\renewcommand{\algorithmicensure}{\textbf{Output:}}
\REQUIRE 5G Traffic Stream $W$, Specialist Ensemble $\{\mathcal{E}_i\}$, drift threshold $\theta$
\ENSURE Optimized Orchestration Policy $\pi^*$
\STATE \textbf{Initialize:} Q-table $Q(s, a) \leftarrow 0$, Learning rate $\alpha$, Discount factor $\gamma$
\FOR{each training episode $m \in \{1,\dots,M\}$}
    \FOR{each discrete decision-making point $t \in \{1,\dots,T\}$}
        \STATE Observe system state $\mathbf{s}_t \in \mathcal{S}$
        \STATE Route incoming data to state-matched LSTM specialist $\mathcal{E}_{s_t}$
        \STATE Compute forecast and evaluate nMAE $e_t$
        \STATE Calculate drift magnitude $\Delta_t = e_t - \theta$
        \STATE Select action $a_t \in \{0, 1\}$ using $\epsilon$-greedy strategy
        \IF{$a_t = 1$}
            \STATE Update weights of $\mathcal{E}_{s_t}$ to restore accuracy
            \IF{$\Delta_t > 0$}
                \STATE $R_t(s_t, a_t) \leftarrow w_1 |\Delta_t| + C_1$ 
            \ELSE
                \STATE $R_t(s_t, a_t) \leftarrow -w_3 |\Delta_t| - C_3$
            \ENDIF
        \ELSE
            \IF{$\Delta_t \le 0$}
                \STATE $R_t(s_t, a_t) \leftarrow w_2 |\Delta_t| + C_2$ 
            \ELSE
                \STATE $R_t(s_t, a_t) \leftarrow -w_4 |\Delta_t| - C_4$
            \ENDIF
        \ENDIF
        \STATE Observe subsequent system state $s'$
        \STATE \textbf{Update Q-value:} $Q(s_t, a_t) \leftarrow Q(s_t, a_t) + \alpha \left[ R_t + \gamma \max_{a'} Q(s', a') - Q(s_t, a_t) \right]$
    \ENDFOR
\ENDFOR
\end{algorithmic}
\label{alg:ortho}
\end{algorithm}

\subsection{Proposed Algorithm}

Algorithm~\ref{alg:rl_drift} takes the incoming user traffic stream $W$, the specialist LSTM ensemble ${\mathcal{E}_i}$, and the drift threshold $\theta$ as inputs and outputs an optimized drift orchestration policy $\pi^*$.
The proposed approach processes the incoming traffic stream in a sequential manner and, at each decision step, the proposed approach observes the system state $\mathbf{s_t}$ based on discretized statistical features such as mean and variance (see line 4), which capture the current operating condition of the network.
Then a state-matched LSTM specialist $\mathcal{E}{s_t}$ is chosen from the ensemble to generate predictions for the current traffic window and evaluates the prediction accuracy by computing the nMAE $e_t$ (see lines 5–6).
Afterwards, the drift magnitude $\Delta_t = e_t - \theta$ is computed to quantify the deviation of the model performance from the acceptable SLA boundary (see line 7), where a positive value indicates the occurrence of drift.

The RL agent selects an action $a_t \in {0,1}$ using an $\epsilon$-greedy strategy (see line 8), where the strategy balances exploration and exploitation to improve policy learning over time.
The action $a_t=1$ triggers retraining of the selected LSTM model to restore predictive accuracy, whereas the action $a_t=0$ skips retraining to preserve computational resources.
When the agent selects the retraining action, the proposed approach updates the weights of the corresponding LSTM specialist using the most recent traffic data (see lines 9–13), enabling the model to adapt to the new user data traffic.
The proposed approach computes the reward $R_t$ based on both the selected action and the drift magnitude (see lines 10–18), where the reward function assigns positive rewards to correct decisions, such as retraining under drift or skipping retraining under stable conditions, and assigns penalties to incorrect decisions, such as unnecessary retraining or failure to retrain during drift. 
After performing the action, the proposed approach observes the subsequent system state $s'$ and updates the Q-value using the Bellman update rule (lines 19–20), enabling the agent to iteratively refine action-value estimates and improve decision-making.

Through repeated interaction with dynamic and varying traffic conditions, the learning process enables the agent to converge toward an optimal policy with the long-term benefit of improved prediction accuracy outweighs the associated computational cost.
The proposed adaptive decision-making approach ensures efficient resource utilization while maintaining reliable AI/ML model performance in B5G networks.

\section{Performance Evaluation}

This section presents the experimental setup used to evaluate the proposed approach.
The evaluation aims to assess how effectively the proposed approach manages drift in real-time traffic scenarios while balancing predictive accuracy and computational cost.
The proposed approach employs a specialist LSTM ensemble to address catastrophic forgetting observed in single-model approaches.
The ensemble consists of multiple state-matched models, each capturing specific traffic patterns, enabling the system to adapt to new traffic conditions while retaining previously learned knowledge. 
Thus, improves robustness to dynamic traffic changes and enhancing predictive performance under dynamic network conditions.

\begin{figure}[!htb]
    \centering
    \vspace{-0.2cm}
    \includegraphics[width=1.0\columnwidth]{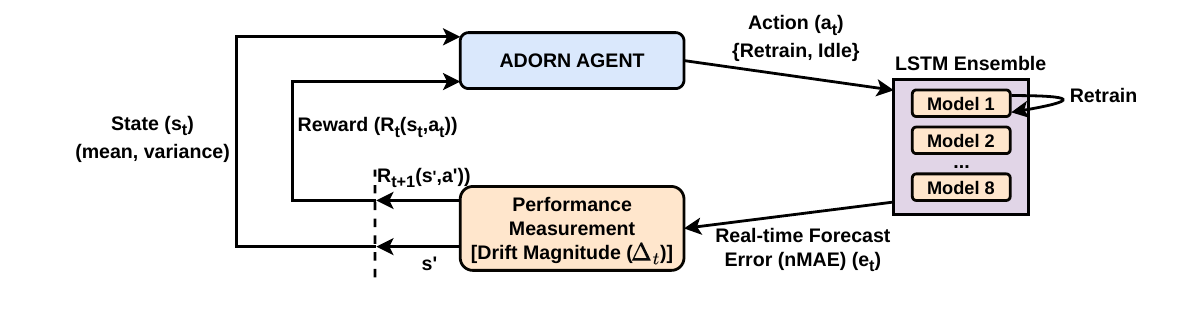}
    \caption{Proposed Q-Learning based ADORN approach}
    \label{fig2a}
    \vspace{-0.1cm}
\end{figure}

The experiment is conducted across eight distinct traffic scenarios generated from the Colosseum traffic dataset~\cite{polese2022colo}, which provides realistic and time-varying network conditions. 
Each traffic scenario exhibits significantly different mean and variance characteristics, thereby defining the system's state space.
Each time step in the dataset includes features --- the number of active users, downlink (DL) traffic rate, and uplink (UL) traffic rate. 
The proposed approach selects the DL traffic rate as the primary prediction target, as it directly affects user QoS and plays a critical role in O-RAN resource management.
The proposed approach pre-processes the dataset by selecting DL traffic corresponding to specific user counts and time steps to capture meaningful traffic variations. Figure.~\ref{fig2a} illustrates the experimental workflow of the proposed Q-Learning-based ADORN approach.

The RL environment defines state space based on statistical traffic features --- mean and variance. Specifically, the continuous state space is discretized by rounding, mapping the environment into 8 distinct states.
The action space consists of two actions: retrain and idle.
The reward function guides the agent to learn an optimal trade-off between prediction accuracy and cumulative retraining actions. 
Table~\ref{tab:rl_params} summarizes the Q-learning hyperparameters, and Table~\ref{tab:lstm_params} summarizes the LSTM ensemble configuration. 
The agent trains over 300 episodes, each consisting of sequential observations of traffic conditions. 
The proposed approach randomly shuffles traffic segments at the beginning of each episode to eliminate temporal bias and ensure that the learned policy depends only on statistical state representation rather than sequence ordering, which improves generalization and robustness under varying traffic conditions.

\begin{table}[!htb]
\centering
\caption{Q-Learning Hyperparameters}
\vspace{-0.2cm}
\label{tab:rl_params}
\begin{tabularx}{\columnwidth}{|Y|Y|} 
\hline
\textbf{Parameter} & \textbf{Value} \\ \hline
Learning Rate ($\alpha$) & 0.1 \\ \hline
Discount Factor ($\gamma$) & 0.9 \\ \hline
Initial Exploration ($\epsilon$) & 1.0 \\ \hline
Epsilon Decay Rate & 0.95 \\ \hline
Drift Threshold ($\theta$) & 0.25 \\ \hline
Total Episodes & 300 \\ \hline
Weighing Coefficient ($w_1, w_2, w_3, w_4$) & 300, 100, 200, 500 \\ \hline
Operational Constants ($C_1, C_2, C_3, C_4$) & 50, 20, 50, 50 \\ \hline
\end{tabularx}
\end{table}

\begin{table}[!htb]
\centering
\caption{LSTM Ensemble Hyperparameters}
\vspace{-0.2cm}
\label{tab:lstm_params}
\begin{tabularx}{\columnwidth}{|Y|Y|}
\hline
\textbf{Parameter} & \textbf{Value} \\ \hline
Sequence Length & 5 \\ \hline
Hidden Units & 64 \\ \hline
Number of Layers & 2 \\ \hline
Dropout Rate & 0.2 \\ \hline
Training Epochs & 150 \\ \hline
Optimizer & Adam \\ \hline
Activation Function & ReLU \\ \hline
\end{tabularx}
\end{table}

To evaluate the effectiveness of the proposed approach, two baseline policies that represent extreme operational strategies --- a Greedy Baseline and a Random Baseline --- are considered for comparison.
\begin{itemize}
\item \textbf{Greedy Baseline:} Prioritizes prediction accuracy without considering the number of retraining actions.
The policy selects the retraining action at every decision step, which minimizes drift impact but incurs high computational overhead.

\item \textbf{Random Baseline:} Selects actions randomly based on a uniform distribution.
The policy does not incorporate system state or traffic dynamics and therefore lacks adaptive decision-making capability.
\end{itemize}

To assess the performance of the proposed approach, three key performance metrics are considered, which are as follows:
\begin{itemize}
\item \textbf{Normalized Mean Absolute Error (nMAE):} It measures forecasting accuracy by quantifying the difference between predicted and actual traffic values.
$$\text{nMAE} = \frac{1}{N} \sum_{t=1}^{N} \frac{|y_t - \hat{y}_t|}{\bar{y}}$$
where, $y_t$ denotes the actual traffic value at time step $t$, $\hat{y}_t$ denotes the predicted value, $\bar{y}$ denotes the mean of the actual values, $N$ is the total number of samples. \emph{nMAE} also serves as an indicator for drift detection based on the pre-defined 25\% threshold~\cite{10802879}.
\item \textbf{Cumulative Reward:} It evaluates the effectiveness of the learned policy by aggregating rewards over time, which reflects how well the proposed approach balances prediction accuracy and cumulative retraining actions.
\item \textbf{Cumulative Retraining Actions:} It quantifies resource efficiency by counting the number of retraining actions, which highlights the ability of the proposed approach to reduce computational and operational overhead compared to baseline approaches.
\end{itemize}

\subsection{Results}

\begin{figure}[!htb]
    \centering
    \vspace{-0.4cm}
    \includegraphics[width=1.0\columnwidth]{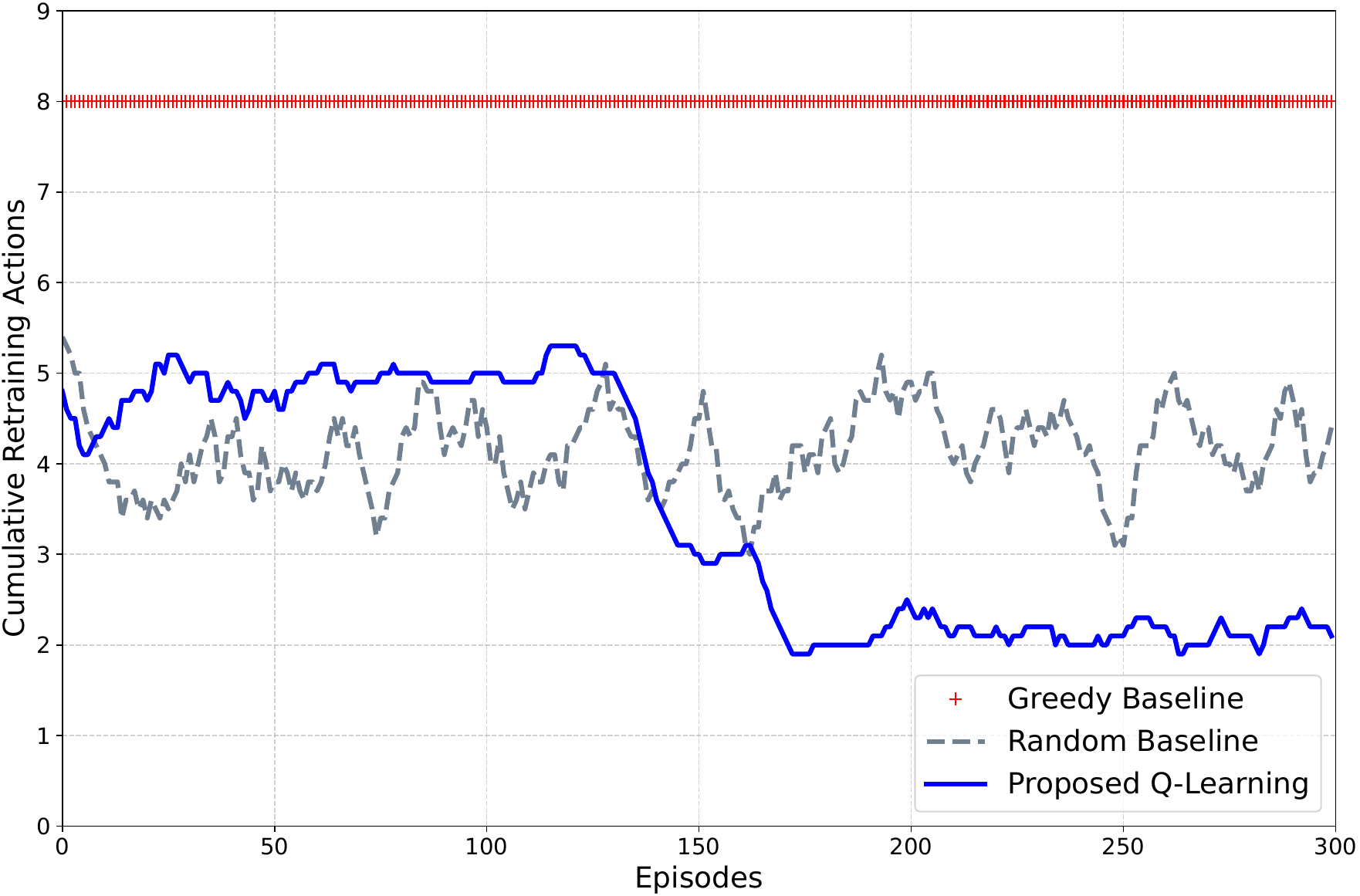}
    \caption{Cumulative Retraining Actions vs Episodes}
    \label{fig3}
    \vspace{-0.2cm}
\end{figure}

Figure.~\ref{fig3} depicts the cumulative retraining actions of the considered approaches across 300 training episodes. 
The curves are smoothed using a moving average with a window size of 10 episodes. 
The greedy baseline maintains a constant 8 retraining actions per episode, as the policy always selects the retraining action without considering computational overhead. 
The random baseline exhibits highly fluctuating behavior, with the smoothed cumulative retraining actions varying approximately 3.0 and 5.4 actions per episode (averaging 4.1 actions), due to stochastic decision-making without awareness of traffic dynamics. 
The proposed Q-learning-based approach demonstrates a consistent learning trend across episodes, with the number of retraining actions gradually decreasing from approximately 4–6 in the initial episodes to about 1–3 in the later episodes, indicating that the agent learns to avoid unnecessary retraining while maintaining predictive performance.

\begin{figure}[!htb]
    \centering
    \vspace{-0.6cm}
    \includegraphics[width=1.0\columnwidth]{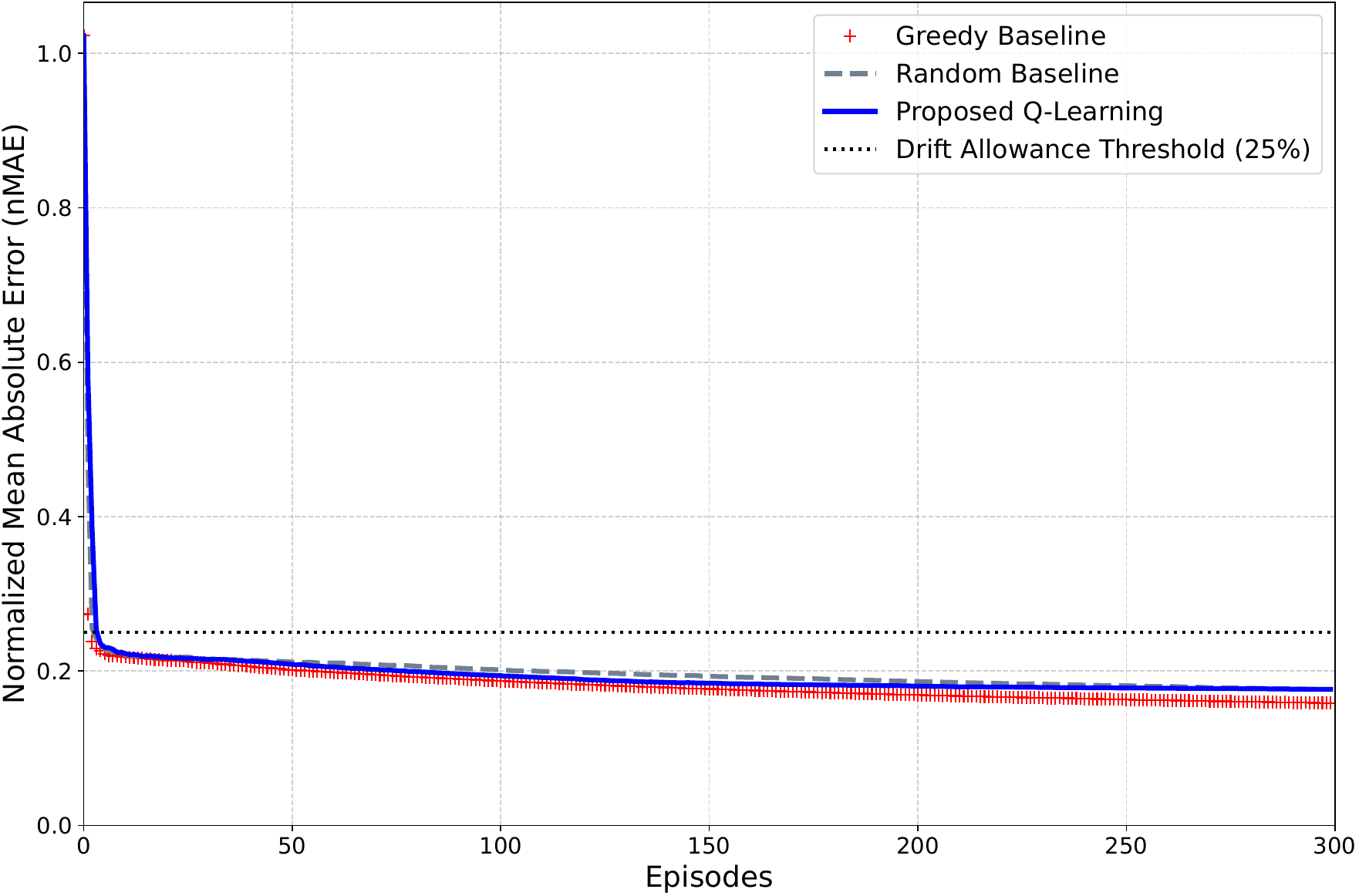}
    \caption{Normalized Mean Absolute Error vs Episodes}
    \label{fig4}
    \vspace{-0.1cm}
\end{figure}

Figure.~\ref{fig4} presents the nMAE of the considered approaches across 300 training episodes, along with the predefined drift tolerance threshold. 
The curves are smoothed using a moving average with a window size of 10 episodes. 
The Greedy baseline achieves the lowest nMAE, gradually decreasing to approximately $0.15$--$0.16$ ($15\%$--$16\%$) by the end of training due to continuous retraining at every step. 
The Random baseline stabilizes around $0.18$--$0.19$ ($18\%$--$19\%$), but lacking stability due to stochastic decision-making. 
The proposed Q-learning-based approach converges to a stable nMAE of approximately $0.17$--$0.18$ ($17\%$--$18\%$), remaining consistently below the drift threshold throughout training. 
Although the Greedy baseline achieves better accuracy, the proposed approach maintains comparable predictive performance while significantly reducing retraining cost, as observed in Fig.~\ref{fig3}.

\begin{figure}
    \centering
    \vspace{-0.5cm}
    \includegraphics[width=1.0\columnwidth]{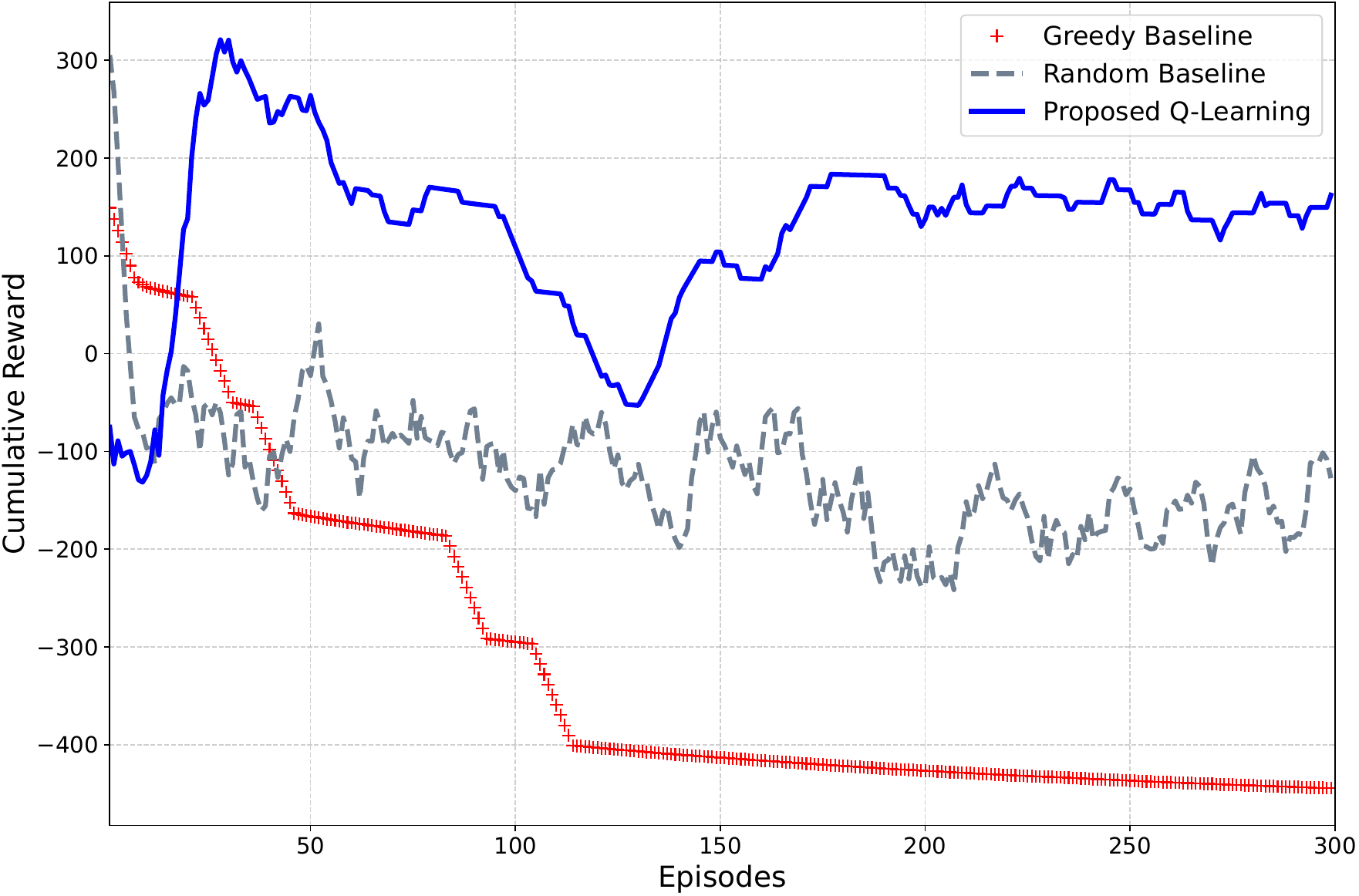}
    \caption{Cumulative Reward vs Episodes}
    \label{fig5}
    \vspace{-0.5cm}
\end{figure}

Figure.~\ref{fig5} depicts the cumulative reward as a function of the number of training episodes. 
During the early exploration phase (i.e., episodes $1$--$15$), the proposed approach exhibits negative rewards, fluctuating between approximately $-132$ and $-17$. 
From around episode $20$, the reward consistently remains positive, fluctuating between $-5$ and $321$, which indicates that the agent begins to learn effective retraining policies.
Beyond episode $150$, the reward stabilizes further, remaining within the range of $80$ to $184$, demonstrating policy convergence.
The Greedy baseline accumulates consistently negative rewards due to redundant retraining actions that incur penalties from unnecessary computational cost.
The Random baseline exhibits highly unstable reward behavior, with frequent penalties caused by both false alarms and unnecessary retraining decisions. 
The proposed approach effectively learns a balanced retraining policy, outperforming the baseline methods in terms of both decision efficiency and long-term reward.

\begin{figure}[htb]
    \centering
    \vspace{-0.25cm}
    \includegraphics[width=1.0\columnwidth]{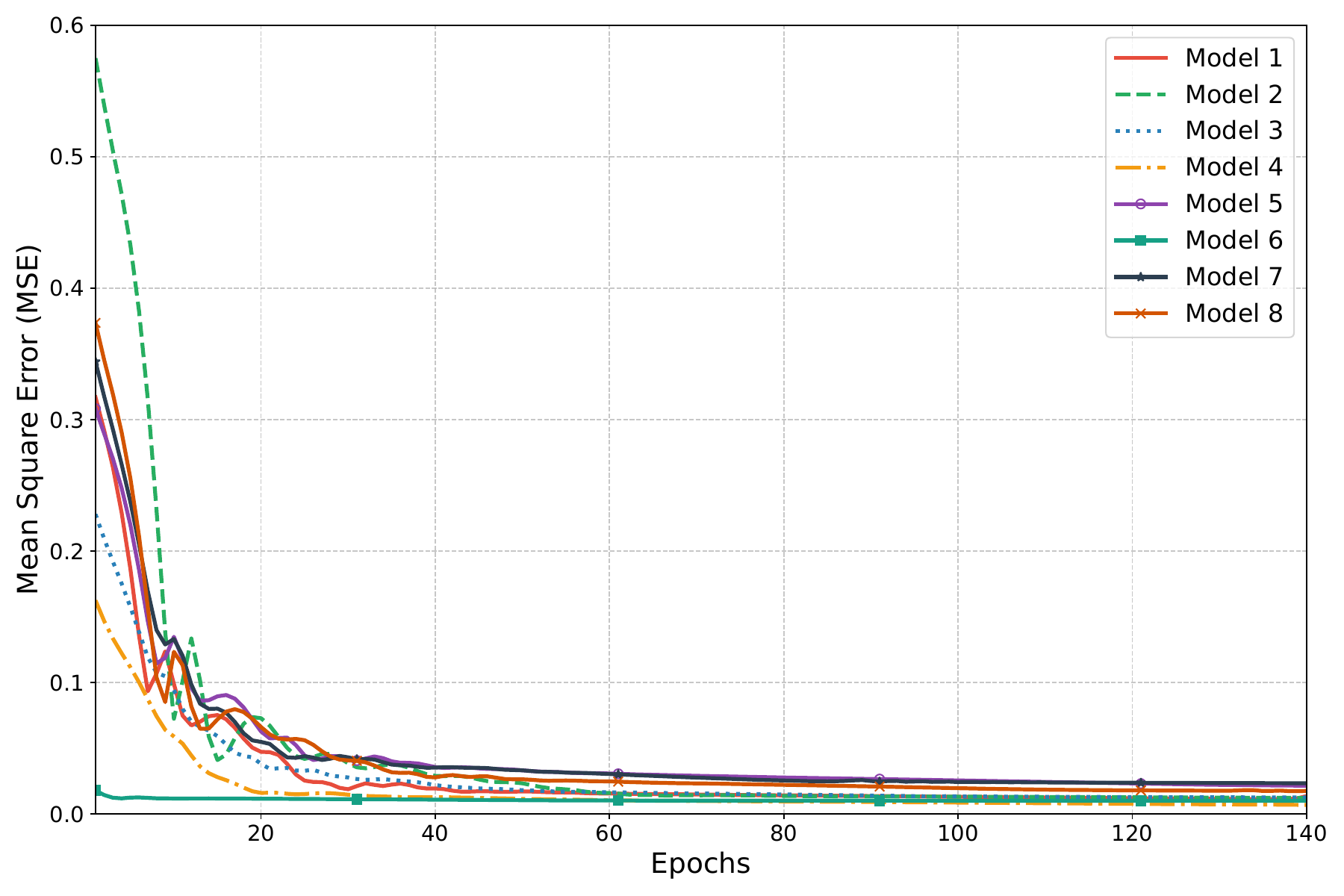}
    \caption{Mean Square Error vs Epochs}
    \label{fig6}
    \vspace{-0.1cm}
\end{figure}

Figure.~\ref{fig6} illustrates the effectiveness of the proposed multi-expert ensemble architecture by presenting the Mean Squared Error (MSE) of eight specialized LSTM models over 140 training epochs. 
Each model corresponds to a specific traffic state. 
The results show a rapid decline in MSE during the initial 10--20 epochs, where most models reduce their initial errors from approximately $0.25$--$0.67$ to below $0.1$, indicating fast convergence during early training. 
As training progresses, all models continue to converge smoothly, reaching stable MSE values between $0.01$ and $0.02$ after approximately 60--100 epochs. 
All eight models exhibit consistent convergence behavior. 
The proposed approach effectively mitigates catastrophic forgetting by assigning each model to a specific traffic state, instead of relying on a single generalized model.

\section{Conclusions and Future Work}

The proposed work introduces ADORN, a Q-learning-based adaptive drift handling approach for O-RAN that formulates the retraining decision as a sequential decision-making problem, enabling intelligent, selective model updates based on real-time system conditions.
Experimental results demonstrate an effective balance between forecasting accuracy and computational efficiency. 
Moreover, the observed reward convergence confirms that the Q-learning agent learns a stable and efficient policy under non-stationary traffic conditions.
The discretized state space restricts scalability in highly dynamic environments with continuous traffic variations, and the use of tabular Q-learning limits generalization to unseen states.
Future work will address these limitations by exploring Deep Reinforcement Learning (DRL) to enable learning over large continuous state spaces for drift management in large-scale O-RAN deployments. 

\section{Acknowledgement}
This work has been partially supported by TTDF “SMARTRIC6G: Smart Drift-Handling Enabler for RAN Intelligent Controllers in 6G Networks (TTDF/6G/422)” project.


\bibliographystyle{IEEEtran}
\bibliography{biblio-ants-24}

\end{document}